\documentclass[prl,aps,twocolumn,showpacs,tightenlines,nofootinbib,preprintnumbers]{revtex4}
\usepackage{graphicx}

\begin{document}

\newcommand{\gsim}{ \mathop{}_{\textstyle \sim}^{\textstyle >} }
\newcommand{\lsim}{ \mathop{}_{\textstyle \sim}^{\textstyle <} }
\newcommand{\vev}[1]{ \left\langle {#1} \right\rangle }

\newcommand{\bear}{\begin{array}}  \newcommand{\eear}{\end{array}}
\newcommand{\bea}{\begin{eqnarray}}  \newcommand{\eea}{\end{eqnarray}}
\newcommand{\beq}{\begin{equation}}  \newcommand{\eeq}{\end{equation}}
\newcommand{\bef}{\begin{figure}}  \newcommand{\eef}{\end{figure}}
\newcommand{\bec}{\begin{center}}  \newcommand{\eec}{\end{center}}
\newcommand{\non}{\nonumber}  \newcommand{\eqn}[1]{\beq {#1}\eeq}
\newcommand{\la}{\left\langle} \newcommand{\ra}{\right\rangle}
\def\lrf#1#2{ \left(\frac{#1}{#2}\right)}
\def\lrfp#1#2#3{ \left(\frac{#1}{#2}\right)^{#3}}

%

\renewcommand{\thefootnote}{\alph{footnote}}

\renewcommand{\thefootnote}{\fnsymbol{footnote}}
\preprint{DESY 07-001}
\title{Anomaly-Induced Inflaton Decay and Gravitino-Overproduction Problem}
\renewcommand{\thefootnote}{\alph{footnote}}

\author{Motoi Endo$^{1}$, Fuminobu Takahashi$^{1}$, and T. T. Yanagida$^{2,3}$}

\affiliation{
${}^1$ Deutsches Elektronen Synchrotron DESY, Notkestrasse 85,
  22607 Hamburg, Germany\\
${}^2${\it Department of Physics, University of Tokyo,
     Tokyo 113-0033, Japan}\\
${}^3${\it Research Center for the Early Universe, University of Tokyo,
     Tokyo 113-0033, Japan}
  }

\begin{abstract}
\noindent
We point out that the inflaton spontaneously decays into any gauge
bosons and gauginos via the super-Weyl, K\"ahler and sigma-model
anomalies in supergravity, once
the inflaton acquires a non-vanishing vacuum expectation value. In
particular, in the dynamical supersymmetry breaking scenarios, the
inflaton necessarily decays into the supersymmetry breaking sector, if
the inflaton mass is larger than the dynamical scale.  This
generically causes the overproduction of the gravitinos, which
severely constrains the inflation models.
\end{abstract}


\maketitle

Inflation~\cite{Guth:1980zm} not only solves basic problems in the big
bang cosmology such as the horizon and flatness problems, but also
provides a natural mechanism to generate density fluctuations
necessary to form the present structure of the universe. In fact, the
standard slow-roll inflation predicts almost scale-invariant power
spectrum, which fits the recent cosmic microwave background
data~\cite{Spergel:2006hy} quite well. During inflation, the universe
is dominated by the potential energy of an inflaton field, and it
expands exponentially~\cite{Sato:1980yn,Guth:1980zm}.  
After inflation, the inflaton transfers its
energy to a thermal plasma by the decay and reheats the universe.  It
is of great importance to unravel the reheating processes to have a
successful thermal history after inflation. Indeed, the reheating is
subject to several constraints; the reheating temperature should be
high enough to generate the baryon asymmetry, while low enough to
avoid the overproduction of unwanted relics.

One usually introduces couplings of the inflaton to the standard-model
particles to cause its decay and hence reheating. The stronger
couplings result in the higher reheating temperature, and so, the
couplings must be so weak to evade the overproduction of the unwanted
relics.  In supergravity, for instance, gravitinos are overproduced by
particle scatterings in thermal plasma, if the reheating temperature
is too high~\cite{Weinberg:1982zq}.  So far, it has been considered
that one can avoid the cosmological difficulties associated with the
unwanted relics (e.g. gravitinos), by setting the coupling of the
inflaton to the visible sector weak enough\footnote{
It should be noted, however, that
Refs.~\cite{Kawasaki:2006gs,Asaka:2006bv,Endo:2006qk} recently pointed
out that the inflaton can decay into the gravitinos, which puts severe
constraints on both the inflation models and the supersymmetry
breaking scenarios.
}.

In this letter we show that, once the inflaton acquires a finite
vacuum expectation value (VEV), it spontaneously decays into any gauge
bosons and gauginos via the quantum effects, anomalies in
supergravity.  Among the super-Weyl, K\"ahler, and sigma-model
anomalies~\cite{Buchbinder:1988yu,LopesCardoso:1992yd}, 
we will concentrate on the effect of the super-Weyl anomaly, for simplicity.
The other K\"ahler and sigma-model anomalies
can affect the inflaton decay at the same order of magnitude\footnote
{
Counter terms~\cite{Bagger:1999rd}  may also contribute to the inflaton decay.
}. 
However, the following
discussion is essentially unchanged even if these are included\footnote
{
The total effect depends on the form of the K\"ahler potential.
For instance, in the case of the minimal K\"ahler potential,
the decay rate is proportional to $(T_G - T_R)^2$ instead of $b_0^2$.
In the case of the K\"ahler potential of the sequestered type,
the contributions from the K\"ahler and sigma model anomalies 
cancel, and the decay is dominantly induced by the
super-Weyl anomaly. 
}.

In the dynamical supersymmetry (SUSY) breaking (DSB)
scenarios, SUSY is broken as a result of strong dynamics in a gauge
theory. The couplings induced by the super-Weyl anomaly makes it
unavoidable for the inflaton to decay into the hidden gauge bosons and
gauginos, which subsequently produce gravitinos. 
The gravitino
production turns out to be prevalent in generic DSB models, which
tightly constrains both the inflation models and SUSY breaking
scenarios.  In particular, as we will see, the gravity-mediation
scenario is almost excluded, and high-scale inflation models such as
hybrid~\cite{Copeland:1994vg} and smooth
hybrid~\cite{Lazarides:1995vr} inflation models are severely
constrained.  We stress that the gravitino production from the
inflaton decay is almost unavoidable, and that it cannot be solved by
taking the reheating temperature low enough. On the contrary, the
lower reheating temperature makes the problem even worse.

We recently pointed out that the inflaton decays into matter
fields in the visible and the hidden SUSY-breaking sectors through
supergravity effects, even without direct couplings between them in
the Einstein frame of the supergravity\footnote{ A part of the decays
is thought of as arising from direct couplings in the conformal frame.
}.  We call it as a spontaneous decay~\cite{Endo:2006qk}. (See also
Ref.~\cite{Watanabe:2006ku} for the non-SUSY case.)  Here, we show
that the inflaton decays into any gauge bosons and gauginos via the
super-Weyl anomaly in the supergravity in addition to the spontaneous
decay.

Let us assume that the inflaton does not have any direct couplings to
the gauge sector. Then, the Lagrangian of the gauge multiplets is
invariant under the super-Weyl transformation at the classical level,
and hence the inflaton decay into the gauge sector is prohibited at 
the tree level~\cite{preparation}.
However the symmetry is anomalous at the quantum level.  The anomaly
not only mediates the SUSY-breaking effects to the visible
sector~\cite{AMSB} but also enables the inflaton to couple to the
gauge supermultiplets. By using the superfield description of the
supergravity, the 1PI effective Lagrangian of the super-Weyl anomaly
is~\cite{Buchbinder:1988yu,LopesCardoso:1992yd}
\beq \Delta\mathcal{L} = \frac{g^2 b_0}{64\pi^2} \int
   d^2\Theta\,2\mathcal{E}W^\alpha W_\alpha \frac{1}{\partial^2}
   (\bar\mathcal{D}^2 -8R) {\bar R} + h.c.  
\eeq
in the conformal frame and in the Planck units: $M_P = 1$.  Here $g$
is a gauge coupling constant, $b_0 = 3T_G-T_R$ is the beta function
coefficient, and $W_\alpha$ is a field strength of corresponding gauge
supermultiplet.  A sum over all matter representations is understood.
The chiral density $\mathcal{E}$, the $\Theta$ variable and the
covariant derivative $\mathcal{D}$ are those defined in the
supergravity~\cite{WessBagger}. 
Note that the inflaton linearly
contributes to the R-current as $b_a \sim \frac{i}{2}
(K_\phi\partial_a \phi - K_\phi^* \partial_a \phi^*)$, and the
superspace curvature ${\bar R}$ contains ${\bar R} = -\frac{1}{6} (M^*
+ \Theta^2(-\frac{1}{2} {\cal R}+ ie^m_a\mathcal{D}_m b^a) + \cdots)$,
where $M$ is a auxiliary field of the supergravity multiplet, and
${\cal R}$ is the Einstein curvature scalar.  Also, the combination
$(K_\phi  \phi + K_\phi^* \phi^*)$ appears in the
scalar component of the graviton.
The inflaton field then
couples to the gauge bosons in the Einstein frame as
\begin{eqnarray}
\label{gauge}
    \Delta\mathcal{L} = \frac{g^2 b_0}{192\pi^2} K_\phi
    \frac{\phi}{M_P} \left(\mathcal{F}_{mn} \mathcal{F}^{mn} -i
    \mathcal{F}_{mn} \tilde \mathcal{F}^{mn}\right) + h.c.
\end{eqnarray}
where $\mathcal{F}_{mn}$ is a field strength of the gauge field and
$\tilde \mathcal{F}^{mn} = \epsilon^{mnkl} \mathcal{F}_{kl}/2$.  In
addition, noting that the F-term satisfies the equation of the motion,
$F^i = -e^{K/2} K^{ij^*} (W_j + K_j W)^*$, the inflaton couples to the
gaugino $\lambda$ as
\begin{eqnarray}
\label{gaugino}
    \Delta\mathcal{L} =  \frac{g^2 b_0}{96\pi^2} 
     K_\phi 
    \frac{m_\phi}{M_P} \phi^* \lambda\lambda + h.c.
\end{eqnarray}
in the Einstein frame. Here we assume that the inflaton mass $m_\phi$
is dominated by the supersymmetric mass term, and used
$F^\phi_{\,\,,\phi^*} \simeq -m_\phi$.  Therefore the inflaton field
couples to the gauge sector through the super-Weyl anomaly as long as
the K\"ahler potential contains a linear term of the inflaton, which
is roughly given by the inflaton VEV $\la \phi \ra$\footnote{
The anomaly-induced decay is suppressed if $\langle K_\phi \rangle = 0$,
while the anomaly itself does not vanish in this case.
}.
 The decay rate becomes
\begin{eqnarray}
\label{eq:decay-rate}
    \Gamma(\phi \to gauge) \;\simeq\;
    \frac{N \alpha^2 b_0^2}{4608 \pi^3} 
     |K_\phi|^2 
    \frac{m_\phi^3}{M_P^2},
\end{eqnarray}
where $N$ is the number of the generators of the gauge group, $\alpha$
is defined as $g^2/4 \pi$, and we assume the canonical normalization
of the inflaton and gauge fields. Here we notice that the half of the
decay rate comes from the decay into the two gauge bosons and the
other half from that into the gaugino pair.

It is interesting to compare the anomaly-induced decay to the recently
observed spontaneous decay which occurs at the tree
level~\cite{Endo:2006qk}. It was shown that the inflaton decays into
the matter fields in the visible and/or hidden sectors, if the
inflaton acquires a finite VEV.  The decay proceeds via both the
Yukawa interactions (with 3-body final states) and the mass terms in
the superpotential, even when there are no direct interaction terms in
the Einstein frame. Thus, for a generic K\"ahler potential (including
the minimal one), the inflaton decay may be dominated by the
spontaneous decay via the top Yukawa coupling.  However, the
anomaly-induced decay rate can be comparable to that of the
spontaneous decay, although it arises at the 1-loop
level. This is because the latter rates are suppressed either by the
phase space of the 3-body final states, or by the mass ratio squared
$(M/m_\phi)^2$ as in the case of the decay into the right-handed
(s)neutrinos with a Majorana mass $M$ satisfying $2M < m_\phi$.

In the DSB scenarios, SUSY is spontaneously broken as a result of
non-perturbative dynamics in a gauge theory, in which case the beta
function coefficient is positive, $b_0 > 0$.  The dynamical scale of
the hidden gauge interactions is related to the SUSY breaking scale as
$\Lambda = x \sqrt{m_{3/2} M_P}$, where $m_{3/2}$ denotes the
gravitino mass, and $x \gtrsim 1$ represents a model-dependent
numerical factor. The gauge bosons and gauginos have masses of
$O(\Lambda)$ due to the strong couplings below the DSB scale.
Therefore, when the inflaton mass, $m_\phi$, is larger than the DSB
scale, the inflaton decays into the SUSY breaking sector via the
super-Weyl anomaly\footnote{
In addition, there may exists a decay into a messenger sector, due to
additional gauge groups introduced to mediate the SUSY breaking
effects.
}, since the decay is kinematically allowed and the mass of the hidden
(s)quarks are much smaller than $m_\phi$ at the decay vertex. 

Let us consider how the decay proceeds. First, the inflaton decays into a
pair of the hidden gauge bosons or gauginos, flying away to the
opposite directions.  Then each of them interacts with the hidden
(s)quarks and hadronizes due to the strong coupling, followed by
cascade decays of the heavy hidden hadrons into lighter ones. The
number of the hidden hadrons produced from each jet, which we call
here as the multiplicity $N_H$, depends on the detailed structure of
the hidden sector such as the gauge groups, the number of the matter
multiplets, and the mass spectrum of the hidden hadrons. We expect,
however, that $N_H$ is in the range of $O(1 - 10^2)$. The hidden
hadrons should eventually decay and release their energy into the
visible sector. The gravitinos are likely to be produced in the decays
of the hidden hadrons~\cite{Casalbuoni:1988kv,Dine:2006ii,Endo:2006tf}
as well as in the cascade decay processes in 
jets\footnote{
In particular, this is the case if the SUSY breaking field 
is a bound state of the hidden (s)quarks.
}.  We denote the averaged number of the gravitinos produced per each jet as
$N_g$. Here we assume each hidden hadron produces one gravitino in the
end, and use the relation $N_g \sim N_H$\footnote{
In a class of the gauge-mediation models of SUSY breaking, the
particles in the hidden sector may dominantly decay into the
standard-model particles~\cite{Fujii:2002fv}.
}. The gravitino abundance is therefore\footnote{
If the inflaton spontaneously decay into the hidden sector at the tree
level~\cite{Endo:2006qk}, more gravitinos will be produced.
}
\bea
Y_{3/2} &=& 2 N_g \frac{ \Gamma_{H}}{\Gamma_{\phi}} 
                  \frac{3 T_{rh}}{4 m_\phi},\non\\
              &\simeq& 3 \times 10^{-7} \xi 
                 \lrfp{m_\phi}{10^{12}{\rm\,GeV}}{2} 
                 \lrf{10^6{\rm\,GeV}}{T_{rh}},
\label{eq:grav-abu}
\eea
where $\Gamma_H$ is the partial decay rate into the hidden gauge
sector given by Eq.~(\ref{eq:decay-rate}), and $\Gamma_{\phi}$ denotes
the total decay rate of the inflaton, related to the reheating
temperature as $T_{rh} \equiv (\pi^2 g_*/10)^{-\frac{1}{4}}
\sqrt{\Gamma_\phi M_P}$.  Here $g_* $ counts the relativistic degrees
of freedom, and we have substituted $g_* = 228.75$ in the second
equality of Eq.~(\ref{eq:grav-abu}).  We also defined $\xi \equiv N
\alpha^2 b_0^2 N_g$, where $N$ and $b_0$ depend on the SUSY breaking
scenarios. For instance, in the IYIT model~\cite{Izawa:1996pk} with a
SU(2) gauge group and four doublet chiral superfields, we have $N=3$,
and $b_0 =4$.

It should be noted that the gravitino abundance (\ref{eq:grav-abu}) is
inversely proportional to the reheating
temperature~\cite{Kawasaki:2006gs}.  That is, for the lower reheating
temperature, more gravitinos are produced. This should be contrasted
to the thermally produced gravitinos, whose abundance is proportional
to the reheating temperature.  For the rest of the paper, we regard
the reheating temperature as a free parameter by introducing
appropriate direct couplings of the inflaton to the visible
sector. We will take the maximal value allowed by cosmological
constraints to give the most conservative estimates on the gravitino
abundance.

The inflaton does not decay into the SUSY breaking sector if $m_\phi
\lesssim \Lambda$.  However, the gravitino pair production then
becomes important~\cite{Kawasaki:2006gs}.  The gravitino
pair production rate is~\cite{pair-gravitino}
\bea
\Gamma_{3/2}^{\rm pair} &\simeq&
\frac{\eta}{96 \pi} |\nabla_\phi G_z|^2 \frac{m_\phi^3}{M_P^2}
\label{eq:grav-pair}
\eea
with $\eta = (m_z/m_\phi)^4$ for $m_\phi > m_z$ and $\eta = 1$ for
$m_\phi < m_z$, where $G \equiv K + \ln |W|^2$, and $m_z$ is the mass
of the SUSY breaking field $z$ with non-vanishing F-term.  Also
$\nabla_\phi G_z$ is defined by $\nabla_\phi G_z \equiv G_{\phi z} -
\Gamma^k_{\phi z} G_k$ with the connection $\Gamma_{ij}^k \equiv
G^{k\ell^*} G_{ij\ell^*}$.  The gravitino abundance is then given by
\vspace{-0.4mm}
\bea Y_{3/2} &=& 2 \frac{\Gamma_{3/2}^{\rm pair}}{\Gamma_{\phi}}
\frac{3 T_{rh}}{4 m_\phi} \simeq 7 \times 10^{-11} \eta \lrfp{\la \phi
\ra}{10^{15}{\rm\,GeV}}{2} \non\\ && ~~~~~~~\times
\lrfp{m_\phi}{10^{12}{\rm\,GeV}}{2} \lrf{10^6{\rm\,GeV}}{T_{rh}},
\label{eq:grav-abu-pair}
\eea
where we have assumed the minimal K\"ahler potential in the last
equality.  Although we have neglected the VEV of $z$ in
(\ref{eq:grav-pair}), including the finite VEV that arises below the
dynamical scale can make the rate even higher~\cite{Endo:2006tf}.
Indeed, taking account of the mixings and couplings between the $\phi$
and $z$ which are radiatively induced for $m_\phi < \Lambda$, the
above gravitino abundance increases. To put it concretely, such an
operator as $|\phi|^2 z z$ in the K\"ahler potential is radiatively
induced, and it additionally contributes to the gravitino production
for $m_\phi > m_z$. The mixings in the K\"ahler potential becomes
important especially for $m_\phi < m_z$.

\begin{figure}[t]
\begin{center}
\includegraphics[width=9cm]{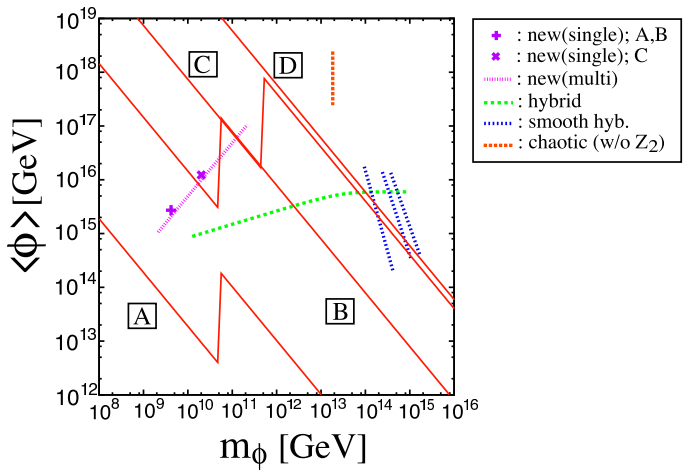}
\caption{Constraints from the gravitino production by the inflaton
decay, for $m_{3/2} = 1{\rm\,TeV}$ with $B_h = 1$ (case\,A), $m_{3/2}
= 1{\rm\,TeV}$ with $B_h = 10^{-3}$ (case\,B), $m_{3/2} =
100{\rm\,TeV}$ (case\,C), and $m_{3/2} = 1{\rm\,GeV}$ (case\,D). The
region above the solid (red) line is excluded for each case. 
 For $m_\phi \gtrsim \Lambda$, we used the anomaly-induced inflaton decay
into the hidden gauge/gauginos to estimate the gravitino abundance
(\ref{eq:grav-abu}), while the gravitino pair production
(\ref{eq:grav-abu-pair}) was used for $m_\phi \lesssim \Lambda$. 
 The typical values of $\la \phi
\ra$ and $m_\phi$ for the single-field new~\cite{Izawa:1996dv},
multi-field new~\cite{Asaka:1999jb}, hybrid~\cite{Copeland:1994vg} and
smooth hybrid~\cite{Lazarides:1995vr}, and
chaotic~\cite{Kawasaki:2000yn} inflation models are also shown.  Note
that we adopt the chaotic inflation model without discrete symmetries,
in which case $\la K_\phi\ra$ is expected to be around the Planck
scale. }
\label{fig:const}
\end{center}
\end{figure}

Using (\ref{eq:grav-abu}) and (\ref{eq:grav-abu-pair}), we can
constrain the inflation models. The results are summarized in
Fig.~\ref{fig:const}, in which we take the maximal value of $T_{rh}$
allowed by the cosmological constraints. For lower $T_{rh}$, the
bounds on $\la \phi \ra$ become severer as $\propto T_{rh}^{1/2}$ for
a fixed $m_\phi$.  We also set $x=1$, $N_g = 10$, $N = 3$, $\alpha =
0.1$, $b_0 = 4$, $\la K_\phi \ra = \la \phi^* \ra$, $m_z= \Lambda$ as
reference values.  We consider the following four cases. In the
cases\,A and B, we set $m_{3/2} = 1{\rm\,TeV}$, assuming the gravitino
is unstable.  The hadronic branching ratio is given by
$B_h=1(10^{-3})$ for the case\,A (B). The gravitino abundance in these
cases are severely constrained by BBN~\cite{Kawasaki:2004yh}.  In the
case\,C, $m_{3/2}=1 {\rm\, GeV}$, and the gravitino is stable. In the
case\,D, we take $m_{3/2}=100{\rm\,TeV}$ with the wino LSP of a mass
given by $2.7 \times 10^{-3} m_{3/2}$~\cite{AMSB}.  The constraints on
$T_{rh}$ and $Y_{3/2}$ come from the requirement that the abundance of
the gravitino (or the winos produced by the gravitino decay) should
not exceed the present dark matter abundance~\cite{Moroi:1993mb}.
From the figure, one can see that the high-scale inflation models such
as the hybrid inflation model are severely constrained, while the new
inflation models may escape the bounds if $B_h$ is suppressed even for
$m_{3/2} = 1{\rm\,TeV}$.

In this letter we have shown that the inflaton decays into any gauge
bosons and gauginos via the super-Weyl anomaly in supergravity, once
the inflaton acquires a nonzero VEV. In particular, the inflaton
necessarily decays into the SUSY breaking sector when the inflaton
mass is larger than the DSB scale. This subsequently produces the
gravitinos, and therefore the gravitino overproduction problem
prevails among the DSB scenarios and most inflation models.

Let us mention that the anomaly-induced decay process and the
associated gravitino problem shown above can be avoided in the
following cases. In the chaotic inflation model with an approximate
$Z_2$ symmetry~\cite{Kawasaki:2000yn,Endo:2006nj}, the VEV of the
inflaton is so suppressed that both the anomaly-induced decay and the
spontaneous decay are suppressed. Similar arguments also apply to
inflation models in the no-scale supergravity~\cite{Endo:2006xg}.

An interesting application of the anomaly-induced inflaton decay can
be found in the case with the K\"ahler potential of the sequestered
type: $K= -3 \ln [1- (|\phi|^2 + |Q|^2)/3]$, where $Q$ collectively
denotes the matter multiplets~\cite{preparation}.  Since there are no
direct couplings of the inflaton to the matter fields in the conformal
frame, the possible decay processes are those mediated by the
supergravity multiplet.  Then, only such an operator that violates the
conformal symmetry induces the inflaton decay, and so, the decay via
the Yukawa couplings does not occur at the tree level.  On the
contrary, the anomaly-induced decay is not suppressed even in this
case. Also, the inflaton decays into the right-handed (s)neutrinos,
since the right-handed Majorana mass violates the conformal
invariance. This may naturally generates the baryon asymmetry via
leptogenesis scenario~\cite{Fukugita:1986hr}.




\end{document}